\documentclass[onecolumn,preprintnumbers,elsart]{revtex4}
\usepackage{eurosym}
\usepackage{makeidx}
\usepackage{amssymb}
\usepackage{amsmath}
\usepackage{mathrsfs}
\usepackage{graphicx}
\usepackage{dcolumn}
\usepackage{bm}
\usepackage[center]{subfigure}
\usepackage{color}

\begin{document}

\title{ Gap and embedded solitons in microwave-coupled binary condensates}
\author{Zhiwei Fan$^{1}$, Zhaopin Chen$^{1}$, Yongyao Li$^{1,2}$ and Boris
A. Malomed$^{1,3}$}
\affiliation{$^{1}$Department of Physical Electronics, School of Electrical Engineering,
Faculty of Engineering, Tel Aviv University, Tel Aviv 69978, Israel\\
$^{2}$School of Physics and Optoelectronic Engineering, Foshan University,
Foshan 528000, China\\
$^{3}$Center for Light-Matter Interaction, Tel Aviv University, Tel Aviv
69978, Israel}

\begin{abstract}
It was recently found that, under the action of the spin-orbit coupling
(SOC) and Zeeman splitting (ZS), binary BEC with intrinsic cubic
nonlinearity supports families of gap solitons, provided that the kinetic
energy is negligible in comparison with the SOC and ZS terms. We demonstrate
that, also under the action of SOC and ZS, a similar setting may be
introduced for BEC with two components representing different atomic states,
resonantly coupled by microwave radiation, while the Poisson's equation
accounts for the feedback of the two-component atomic wave function onto the 
radiation. The microwave-mediated interaction induces an effective nonlinear
trapping potential, which strongly affects the purport of the linear
spectrum in this system. As a result, families of both gap and \textit{%
embedded} solitons (those overlapping with the continuous spectrum) are
found, which are chiefly stable. The shape of the solitons features exact or
broken skew symmetry. In addition to fundamental solitons (whose shape may
or may not include a node), a family of dipole solitons is constructed too,
which are even more stable than their fundamental counterparts. A nontrivial
stability area is identified for moving solitons in the present system,
which lacks Galilean invariance. Colliding solitons merge into a single one.
\end{abstract}

\maketitle

\section{Introduction}

The realization of the spin-orbit coupling (SOC)\ in quantum atomic gases
\cite{SOC1}-\cite{SOC-review2} and exciton-polariton condensates in
semiconductor microcavities \cite{microcav1}-\cite{microcav5} has initiated
a new direction in experimental and theoretical studies of atomic and
photonic waves. While SOC is a linear effect, its interplay with intrinsic
nonlinearity of Bose-Einstein condensates (BECs) makes it possible to
predict the creation of topological modes in these settings, such as
vortices \cite{SOC-vortices}, monopoles \cite{SOC-monopole}, and skyrmions
\cite{SOC-skyrmion,SOC-skyrmion2}, as well as stable one-dimensional \cite%
{Brand}-\cite{soliton-1D-6}, two-dimensional \cite{Ben Li}-\cite{HS-Rabi},
and three-dimensional\ \cite{Han Pu} solitons, see also a brief review in
Ref. \cite{EPL}.

A specific approach to the creation of stable one- and two-dimensional
solitons in two-component BEC subject to the combined action of SOC and
Zeeman splitting (ZS), which is a generic ingredient of SOC systems \cite%
{Zeeman}, was recently proposed in Refs. \cite{we} and \cite{HS}, under the
condition that the SOC and ZS terms in the system's Hamiltonian are much
larger than the kinetic energy. Neglecting, accordingly, the second
derivatives in the respective system of coupled Gross-Pitaevskii equations
(GPEs), which include the combination of the SOC and ZS terms, one obtains a
linear spectrum with a finite bandgap. The cubic nonlinearity induced by
atomic collisions in BEC may readily create families of solitons populating
the bandgap. In this connection, it is relevant to mention that gap solitons
were predicted \cite{Sterke,Sterke2,Sterke3} and experimentally demonstrated
\cite{Krug,BJE} in fiber-optic Bragg gratings, in polariton condensates
under the action of photonic lattices \cite{plasmon,plasmon2}, and in the
single-component BEC\ loaded in an optical-lattice potential \cite%
{BEC,BEC2,Oberthaler}.

The consideration of such models (they may be loosely defined as those for
\textquotedblleft heavy atoms", whose kinetic energy may be omitted) is
relevant because, as it was shown in Ref. \cite{we}, if a low-dimensional
SOC\ system is derived from the three-dimensional one, being subject to the
action of tight confinement in the transverse direction(s), the kinetic
energy in the ensuing system of GPEs is indeed much smaller than the
energies corresponding to the SOC terms. The possibility to omit the terms
with the second derivatives also plays a crucial role in the context of BEC
systems with a flatband spectrum \cite{flatband}.

Another relevant setting for the interplay of SOC and nonlinearity is
offered by the system composed of hyperfine atomic states resonantly coupled
by the magnetic component of the microwave radiation, with the feedback of
the atomic states on the radiation governed by the corresponding Poisson's
equation \cite{Jieli1}. This system gives rise to hybrid
matter-wave-microwave solitons (somewhat similar, in this respect, to
exciton-polariton solitons in semiconductor microcavities \cite%
{exc-polar-sol,plasmon,plasmon2}). The SOC was recently added to this model
in Ref. \cite{Jieli-SOC}. The aim of the present work is to transform the
system into one dominated by the first-order spatial derivatives, accounting
for the SOC, and produce families of stable solitons in the system. An
essential peculiarity of the setting is that the effectively nonlocal
interaction between the components, mediated by the microwave field,
strongly affects the concept of the linear spectrum, by adding an effective
nonlinear trapping potential to the linearized system, see Eq. (\ref{II})
below. In other words, the nonlocal nonlinearity makes the system \emph{%
non-linearizable}, and alters the fundamental significance of the bandgap.
As a result, the system generates, depending on the relative strength of
contact and microwave-induced nonlinear terms, families of gap solitons, as
well as families of \textit{embedded} solitons \cite{embedd1}-\cite{embedd3}%
, which exist inside the radiation band, while the bandgap remains empty.
Another noteworthy findings is that, on the contrary to what is commonly
known about ground states in linear settings, in the present nonlinear
system fundamental solitons, which realize the ground state, may feature a
node (zero crossing) in their profile. In addition to the fundamental modes,
we also find stable excited states in the form of dipole solitons.
Interestingly, their stability area is \emph{broader} than the one for the
fundamental counterparts. A stability area is also identified for moving
fundamental solitons, which is a nontrivial finding for the present system,
which is not subject to the Galilean invariance. Interaction between
counterpropagating solitons leads to multiple collisions and eventual merger.

The main part of the paper is organized as follows. The model and some
analytical results are formulated in Section II. Numerical findings for
fundamental and dipole solitons, as well as for moving ones, are summarized
in Section III. The paper is concluded by Section IV.

\section{The model}

\begin{figure}[tbp]
\centering{\label{fig1a} \includegraphics[scale=0.15]{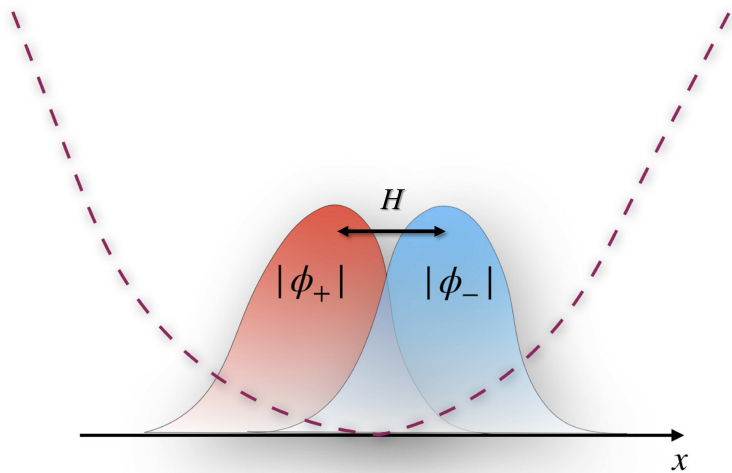}}
\caption{A sketch of the system composed of the binary Bose-Einstein
condensate, with components $(\protect\phi _{+},\protect\phi _{-})$, trapped
in the effective nonlinear potential (shown by the dashed curve) induced by
the microwave-mediated interaction, and linearly coupled by the effective
Rabi mixing $\sim H$, designated by the double arrow ($\leftrightarrow $).
The sketch shows the shape of a typical skew-symmetric soliton (the ZS will
break the mutual symmetry of the two components).}
\label{1a}
\end{figure}

The binary BEC, with two components $\phi _{\pm }\left( x,t\right) $ of the
mean-field wave function coupled linearly by SOC and nonlinearly by the
microwave radiation, is modeled by a system of one-dimensional GPEs. In a
scaled form, the system, in which the kinetic-energy terms are neglected (as
said above), is written, following Ref. \cite{Jieli-SOC} (in which the
kinetic energy was kept), as

\begin{gather}
i\partial _{t}\phi _{+}=\partial _{x}\phi _{-}-\Omega \phi _{+}-H\phi
_{-}-g|\phi _{+}|^{2}\phi _{+}+\frac{\gamma }{2}{\phi }_{-}\int_{-\infty
}^{+\infty }|x-x^{\prime }|{\phi }_{-}^{\ast }(x^{\prime }){\phi }%
_{+}(x^{\prime })dx^{\prime },  \notag \\
i\partial _{t}{\phi }_{-}=-\partial _{x}\phi _{+}+\Omega \phi _{-}-H\phi
_{+}-g|\phi _{-}|^{2}\phi _{-}+\frac{\gamma }{2}{\phi }_{+}\int_{-\infty
}^{+\infty }|x-x^{\prime }|{\phi }_{-}(x^{\prime }){\phi }_{+}^{\ast
}(x^{\prime })dx^{\prime },  \label{basiceq}
\end{gather}%
where $\ast $ stands for the complex-conjugate expression. Here, the
coefficient of SOC, represented by the $x$-derivatives, is scaled to be $1$,
$\Omega $ is the ZS strength, which may be imposed by dc magnetic field (
alternatively, the same terms may be represent the Stark - Lo Surdo effect,
imposed by dc electric field), the amplitude of the magnetic component of
the background microwave field is $H$ (in fact, it induces an effective Rabi
mixing in the system, cf. Refs. \cite%
{Rabi-Luca-me,HS-Rabi,Rabi-Wesley,Rabi-Luca}), and the contact
self-interaction in each component (if any) is represented by coefficient $g$
($g>0$ and $g<0$ correspond to self-attraction and repulsion, respectively).
Contact cross-interaction between the components can be readily added to the
system, in the form of terms $\sim \left\vert \phi _{\mp }\right\vert
^{2}\phi _{\pm }$ in each equation, but they do not produce essential
changes in the results reported below. The integral terms represent the
feedback of the microwave field, generated by transitions between the $\phi
_{\pm }$ components, with strength $\gamma >0$, on these components. By
means of rescaling admitted by Eq. (\ref{basiceq}), we\ set $\gamma =0.5$
and $\Omega =1$, unless $\Omega =0$ (in Ref. \cite{Jieli-SOC}, different
scaling was adopted, with $\gamma =0.02$). The integral terms are actually
generated by the solution of the respective one-dimensional Poisson
equation, written in terms of the corresponding Green's function, $G\left(
x,x^{\prime }\right) =(1/2)\left\vert x-x^{\prime }\right\vert $ \cite%
{Jieli-SOC}. A schematic of the system under the consideration is displayed
in Fig. \ref{1a}.

Stationary states with chemical potential $\mu $ are looked for as solutions
to Eq. (\ref{basiceq}) in the form of
\begin{equation}
\phi _{\pm }=\exp (-i\mu t)u_{\pm }(x),  \label{phiu}
\end{equation}%
with real stationary wave functions $u_{\pm }$ obeying the following
integro-differential equations:

\begin{eqnarray}
\mu u_{+} &=&\partial _{x}u_{-}-\Omega u_{+}-Hu_{-}-gu_{+}^{3}+\frac{\gamma
}{2}u_{-}\int_{-\infty }^{+\infty }|x-x^{\prime }|{u}_{-}(x^{\prime }){u}%
_{+}(x^{\prime })dx^{\prime },  \notag \\
\mu u_{-} &=&-\partial _{x}u_{+}+\Omega u_{-}-Hu_{+}-gu_{-}^{3}+\frac{\gamma
}{2}u_{+}\int_{-\infty }^{+\infty }|x-x^{\prime }|{u}_{-}(x^{\prime }){u}%
_{+}(x^{\prime })dx^{\prime }.  \label{equmu}
\end{eqnarray}%
Note that, in the absence of the ZS, $\Omega =0$, Eq. (\ref{equmu}) admits
the \textit{skew-symmetry }reduction,%
\begin{equation}
u_{+}(x)=\pm u_{-}(-x),  \label{skew}
\end{equation}%
which is broken by the ZS terms.

Equations (\ref{basiceq}) conserve the total norm,%
\begin{equation}
N\equiv \int_{-\infty }^{+\infty }\left(
|u_{+}(x)|^{2}+|u_{-}(x)|^{2}\right) dx,  \label{norm}
\end{equation}%
and energy (Hamiltonian),%
\begin{gather}
E=\int_{-\infty }^{+\infty }dx\left[ \phi _{+}^{\ast }\partial _{x}\phi
_{-}-\phi _{-}^{\ast }\partial _{x}\phi _{+}+\Omega \left( \left\vert \phi
_{-}\right\vert ^{2}-\left\vert \phi _{+}\right\vert ^{2}\right) -H\left(
\phi _{+}^{\ast }\phi _{-}+\phi _{-}^{\ast }\phi _{+}\right) \right.  \notag
\\
\left. -\frac{g}{2}\left( \left\vert \phi _{+}\right\vert ^{4}+\left\vert
\phi _{-}\right\vert ^{4}\right) \right] +\frac{\gamma }{2}\int_{-\infty
}^{+\infty }\int_{-\infty }^{+\infty }\left\vert x-x^{\prime }\right\vert
dxdx^{\prime }\left[ \phi _{+}^{\ast }(x)\phi _{-}(x){\phi }_{-}^{\ast
}(x^{\prime }){\phi }_{+}(x^{\prime })\right] .  \label{E}
\end{gather}%
Note that, with regard to the possibility of the integration by parts,
expression (\ref{E}) is real, even if the SOC terms in the integrand seem
complex.

The linearization of Eq. (\ref{equmu}) produces the dispersion relation in
the free space (in the absence of an external trapping potential),
\begin{equation}
\mu ^{2}=\Omega ^{2}+H^{2}+k^{2},  \label{BGsquare}
\end{equation}%
which demonstrates that the background magnetic field and ZS contribute to
the formation of the spectral gap,
\begin{equation}
-\sqrt{\Omega ^{2}+H^{2}}<\mu <+\sqrt{\Omega ^{2}+H^{2}},  \label{BGgap}
\end{equation}%
in which one may expect the creation of gap solitons under the action of the
system's nonlinearities, cf. Refs. \cite{we} and \cite{HS}. The gap is
located between semi-infinite bands populated by radiation modes.
Nevertheless, it is shown below that, in addition to the in-gap solitons,
the system readily creates families of \textit{embedded solitons} in the
bands, while the gap is left empty. This finding can be readily understood,
as Eq. (\ref{equmu}) takes the following asymptotic form at $|x|\rightarrow
\infty $:%
\begin{eqnarray}
\mu u_{+} &=&\partial _{x}u_{-}-\Omega u_{+}-Hu_{-}+\frac{\gamma }{2}%
I|x|u_{-},  \notag \\
\mu u_{-} &=&-\partial _{x}u_{+}+\Omega u_{-}-Hu_{+}+\frac{\gamma }{2}%
I|x|u_{+}.  \label{II}
\end{eqnarray}%
where%
\begin{equation}
I\equiv \int_{-\infty }^{+\infty }u_{-}(x^{\prime })u_{+}(x^{\prime })dx
\label{I}
\end{equation}%
is a constant. The effective cross-potential in Eq. (\ref{II}), growing $%
\sim |x|$ at $|x|\rightarrow \infty $, completely changes the definition of
the system's spectrum, and may make the distinction between the gap and
bands, predicted by the linearization of the system, irrelevant. Indeed, the
asymptotic form of the solution to Eq. (\ref{II}) is%
\begin{eqnarray}
\left\{ u_{-}(x),u_{+}(x)\right\} &\approx &u_{0}^{(-)}\left\{ 1,\frac{\mu
-\Omega }{\gamma I}x^{-1}\right\} \exp \left( -\frac{\gamma }{4}%
|I|x^{2}\right) ~\mathrm{at}~x\rightarrow +\infty ,  \notag \\
\left\{ u_{+}(x),u_{-}(x)\right\} &\approx &u_{0}^{(+)}\left\{ 1,\frac{\mu
+\Omega }{\gamma I}|x|^{-1}\right\} \exp \left( -\frac{\gamma }{4}%
|I|x^{2}\right) ~\mathrm{at}~x\rightarrow -\infty ,  \label{asympt}
\end{eqnarray}%
where $u_{0}^{(\mp )}$ are constants. In the case of $\Omega =0$, this
solution has $u_{0}^{(-)}=-u_{0}^{(+)}$, satisfying the skew-symmetry
condition (\ref{skew}) with the bottom sign [the particular sign is selected
by comparison with the full numerical solution (\ref{equmu})].

The Gaussian asymptotic form produced by Eq. (\ref{asympt}) is drastically
different from the exponential expressions for tails of ordinary gap
solitons, which explains the possibility of finding soliton families in the
band, rather than in the gap, as shown below by numerically found solutions.
In fact, the asymptotic form (\ref{II}) implies that the underlying GPE
system (\ref{basiceq}) is \emph{non-linearizable}, cf. Ref. \cite{Barcelona}%
. Comparison between numerical solutions and analytical prediction given by
Eq. (\ref{asympt}) is also shown below, in Fig. \ref{11a}.

\section{Numerical results}

\subsection{Fundamental solitons}

\begin{figure}[h]
\centering{\label{fig3a} \includegraphics[scale=0.3]{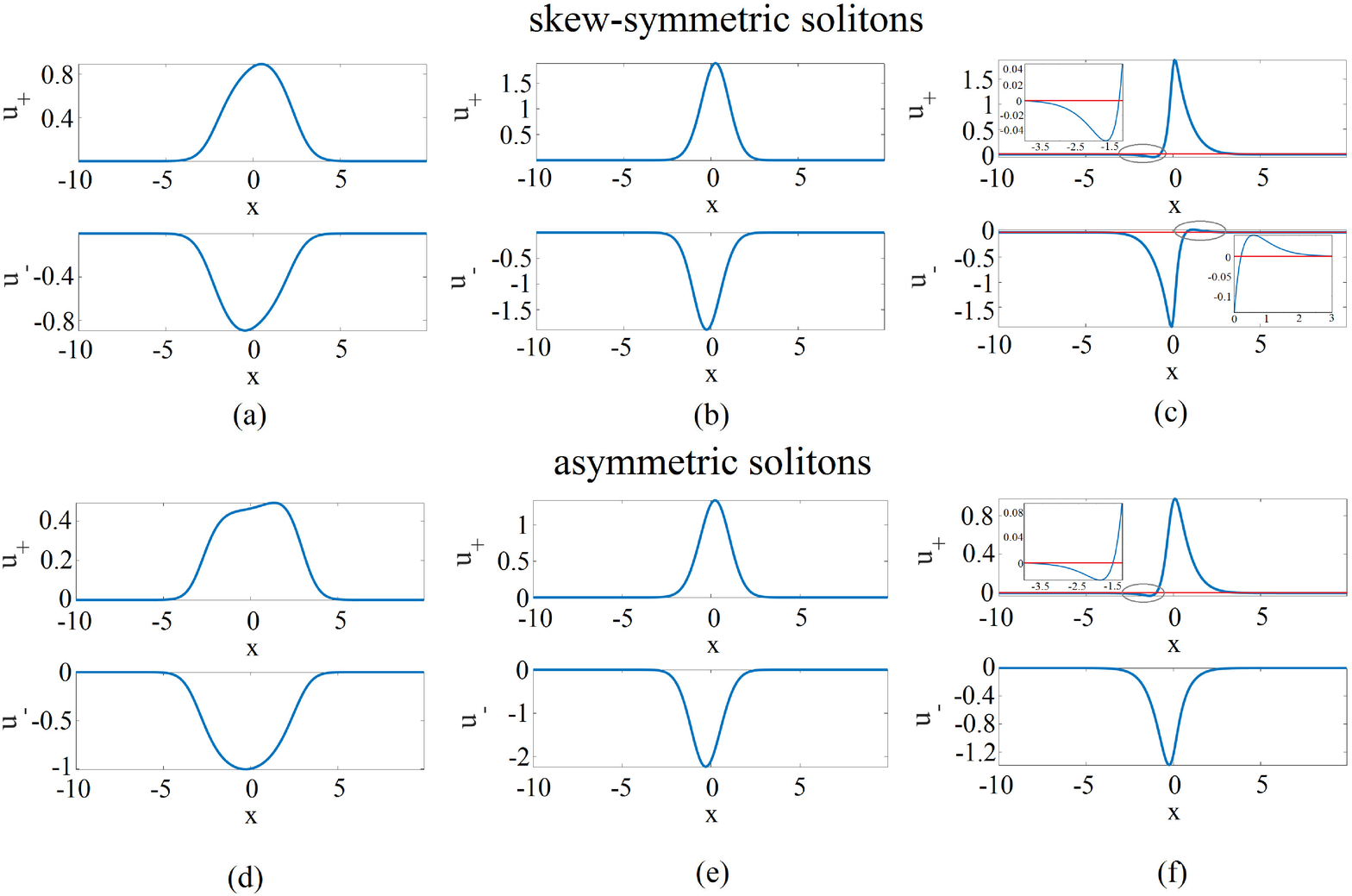}}
\caption{Panels (a,b,c) and (d,e,f) display typical examples of ground-state
soliton profiles which are, respectively, skew-symmetric (with $\Omega =0$)
and asymmetric (with $\Omega =1$). The other parameters are $\protect\gamma %
=0.5$ and (a) $(N,H,g)=(5,1,-1)$, belonging to the magenta line in Fig.
\protect\ref{2a}(a); (b) $(N,H,g)=(10,1,0)$, belonging to the blue line in
Fig. \protect\ref{2a}(a); (c) $(N,H,g)=(5,1,1)$, belonging to the red line
in Fig. \protect\ref{2a}(c); (d) $(N,H,g)=(5,1,-1)$, belonging to the orange
line in Fig. \protect\ref{2a}(d); (e) $(N,H,g)=(10,1,0)$, belonging to the
green line in Fig. \protect\ref{2a}(d); (f) $(N,H,g)=(3,1,1)$, belonging to
the violet line in Fig. \protect\ref{2a}(f). A nontrivial feature, observed
in the in-gap solitons in panels (c) and (f), is that a sufficiently strong
contact self-attraction, with $g=1$, creates\emph{\ zero-crossings} (nodes)
in the ground-state profiles. The nodes are clearly shown in insets to
panels (c) and (f), which zoom the wave-function profiles in regions denoted
by gray circles in the main plots. Other solitons [shown in panels (a), (b),
(d), and (e)] are embedded (in-band) ones, which are free of nodes. The
soliton in (c) is unstable, all other ones being stable.}
\label{3a}
\end{figure}

\begin{figure}[h]
\centering{\label{fig2a} \includegraphics[scale=0.45]{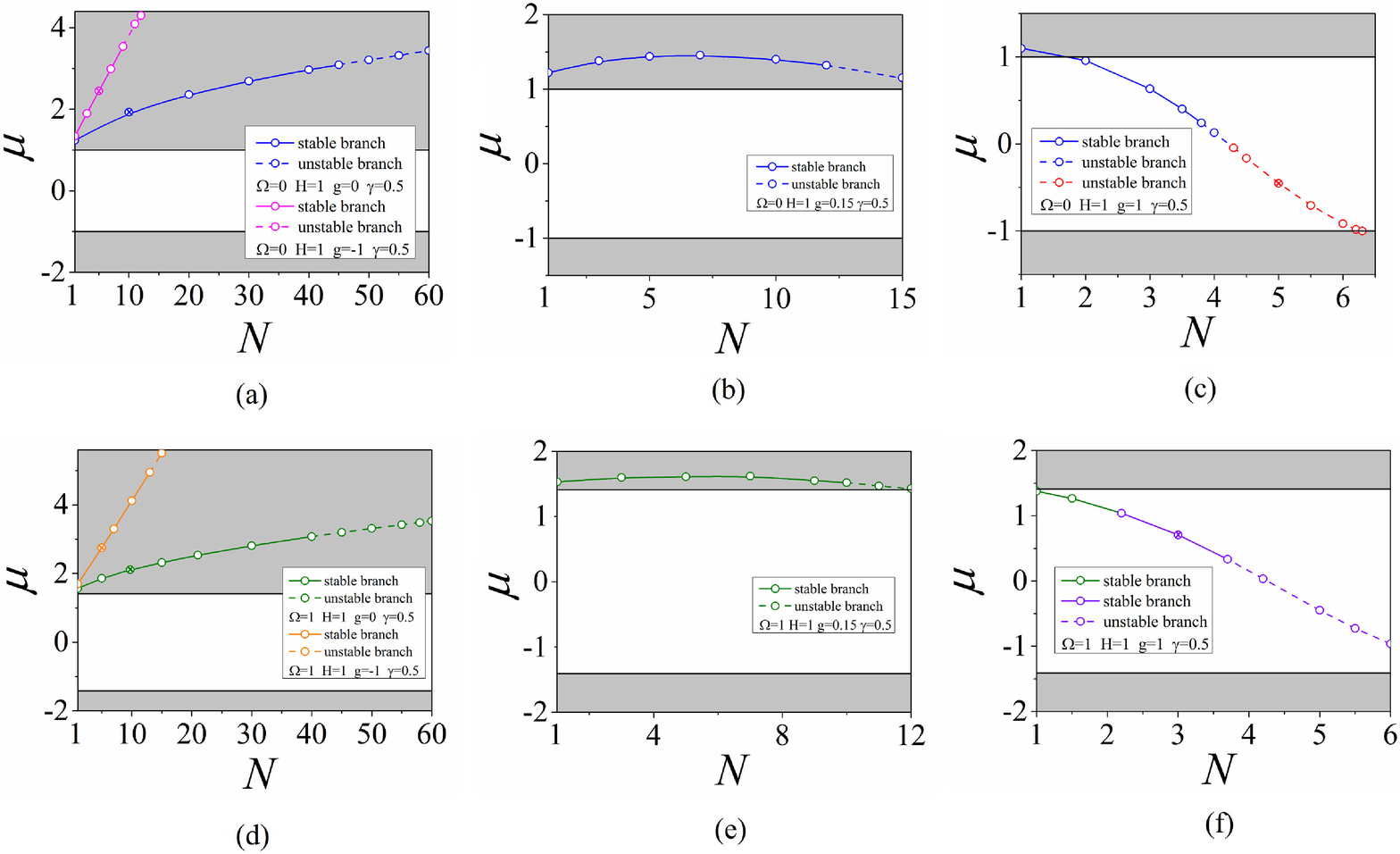}}
\caption{Dependences $N(\protect\mu )$ for six generic soliton families, at
values of parameters indicated in panels. Solid and dashed lines represent
stable and unstable segments of the families, respectively. In panels (c)
and (f), red and violet lines (in the lower right) represent solitons with nodes, while blue and
green ones (in the upper left)  designate families of nodeless solitons. In each panel, white and
gray areas represent, respectively, the bandgap, predicted by Eq. (\protect
\ref{BGgap}) for the linearized system, and the bands above and beneath the
gap. Coordinates of points separating stable and unstable segments, as well
as ones carrying nodeless and noded solitons, in different panels are: (a) $%
(N,\protect\mu )=(9,3.54)$ and $(45,3.09)$; (b) $(N,\protect\mu )=(12,1.32)$%
; (c) $(N,\protect\mu )=(3.8,0.24)$ and $(N,\protect\mu )=(4.3,-0.045)$
(this point separates the blue and red branches); (d) $(N,\protect\mu %
)=(13,4.95)$ and $(N,\protect\mu )=(40,3.08)$; (e) $(N,\protect\mu %
)=(10,1.52)$ ; (f) $(N,\protect\mu )=(3.7,0.33)$ and $(N,\protect\mu %
)=(2.2,1.04)$ (this point separates the green and violet branches). In
panels (b) and (e) points corresponding to the largest values of $\protect%
\mu $ are $(N,\protect\mu )=(7,1.46)$ and $(N,\protect\mu )=(7,1.62)$,
respectively. In panels (a,c,d,f), examples of solitons shown in Fig.
\protect\ref{3a} are marked by cross-in-circle symbols.}
\label{2a}
\end{figure}

Soliton solutions of Eq. (\ref{equmu}) were produced by means of the
squared-operator iteration method \cite{JKY}. Then, their stability or
instability was identified by means of direct simulations of the perturbed
evolution.

First, in Figs. \ref{3a}(a,b,c) and (d,e,f) we display, severally, typical
examples of stationary wave functions $u_{\pm }(x)$ for skew-symmetric and
asymmetric solitons. All these solutions may be identified as ground states.
In particular, the modes shown in panels (a,b,d,e) comply with the
fundamental principle, borrowed from linear theories, which states that
ground-state profiles must not have zero-crossing points (nodes).
Nevertheless, the ground states presented in panels (c) and (f) break this
principle, each featuring one node, which may be possible in nonlinear
systems [in the present case, the nodes emerge if the relatively strong
self-attraction terms, with $g=1$, are present in Eq. \ref{basiceq})]. It is
relevant to note that, in the absence of the background magnetic field ($H=0$%
), the ground state of the SOC system takes the form of the gap soliton,
with one spatially even nodeless component, and the other odd one, which has
the node at the center \cite{HS}. Accordingly, the nodes observed in Figs. %
\ref{3a}(c) and (f) may be considered as \textquotedblleft remnants" of the
nodes in the above-mentioned odd component. The solitons with nodes,
displayed in (c) and (f), are identified as ground states, as they belong to
families which are produced by continuation of ones representing nodeless
ground-state solitons -- for instance, the blue and green lines in Figs. \ref%
{2a}(c) and (f) are obtained as continuous extensions of the red and violet
segments. No nodeless modes could be found for those values of $N$ at which
the noded ground-state solitons have been produced by the solution of Eq. (%
\ref{equmu}).

A natural trend, demonstrated by the comparison of different profiles in
Fig. \ref{3a}, is that the transition from the self-repulsion to attraction,
i.e., from $g<0$ to $g>0$, leads to compression of each component. Further,
the two components generate each other linearly via the SOC and Rabi
coupling (the latter one represented by $H$), and simultaneously they
mutually repel nonlinearly, through the effective cross-potentials $\sim
\gamma $ in Eq. (\ref{II}). Accordingly, the self-compressed components with
larger amplitudes stronger repel each other, featuring effective separation
in Figs. \ref{3a}(c,f). Finally, the strong separation makes coefficient $%
|I| $ (\ref{I}) smaller, which weakens the role of the effective confining
potentials in Eqs. (\ref{II}) and (\ref{asympt}) in comparison with the
linear SOC and Rabi terms in the asymptotic area, $|x|\rightarrow \infty $,
thus allowing the solitons to populate the linearly-predicted gap, as seen
below in Figs. \ref{2a}(c,f).

Systematic results, including the location of families of the skew-symmetric
and asymmetric solitons with respect to the bandgap (\ref{BGgap}), and their
stability, are summarized in Fig. \ref{2a}, by means of $\mu (N)$
dependences for the soliton families in six generic cases. Only in panels
(c) and (f) these are families of in-gap solitons. In other cases they
exist, as embedded solitons, in radiation bands, while the gap remains
empty. Embedded solitons were not found in Ref. \cite{HS}, while families of
gap solitons obtained in the model with the dipole-dipole interaction,
considered in Ref. \cite{we}, extend into the bands (however, gaps were
never empty in that system).

It is worthy to note that the Vakhitov-Kolokolov (VK)\ \cite{VK1,VK2,JKY} or
anti-VK \cite{anti} criteria, which relate the sign of the slope, $d\mu
/dN<0 $ or $d\mu /dN>0$, to the necessary stability condition for solitons
which are supported, severally, by the self-attractive or repulsive
nonlinearities, is valid in the present system when both the nonlocal and
local nonlinearities are self-repulsive. Indeed, in Figs. \ref{2a}(a,d), the
families satisfy the anti-VK criterion, $d\mu /dN>0$, and are indeed
completely stable. On the other hand, it is not surprising that, in the case
when the solitons are supported by the combination of the nonlocal repulsion
and contact attraction, the VK/anti-VK criterion does not hold, as it is not
possible to identify the dominant nonlinear term: in Figs. \ref{2a}(b,e),
the change of the sign from $d\mu /dN>0$ to $d\mu /dN<0$ does not lead to
destabilization of the solitons (non-compliance with the VK criterion occurs
in other models too \cite{JKY}). As concerns the evolution of unstable
solitons and interaction between stable ones, they are similar to examples
displayed below in Figs. \ref{6a}(b) and (c).

\begin{figure}[tbp]
\centering{\label{fig4a} \includegraphics[scale=0.6]{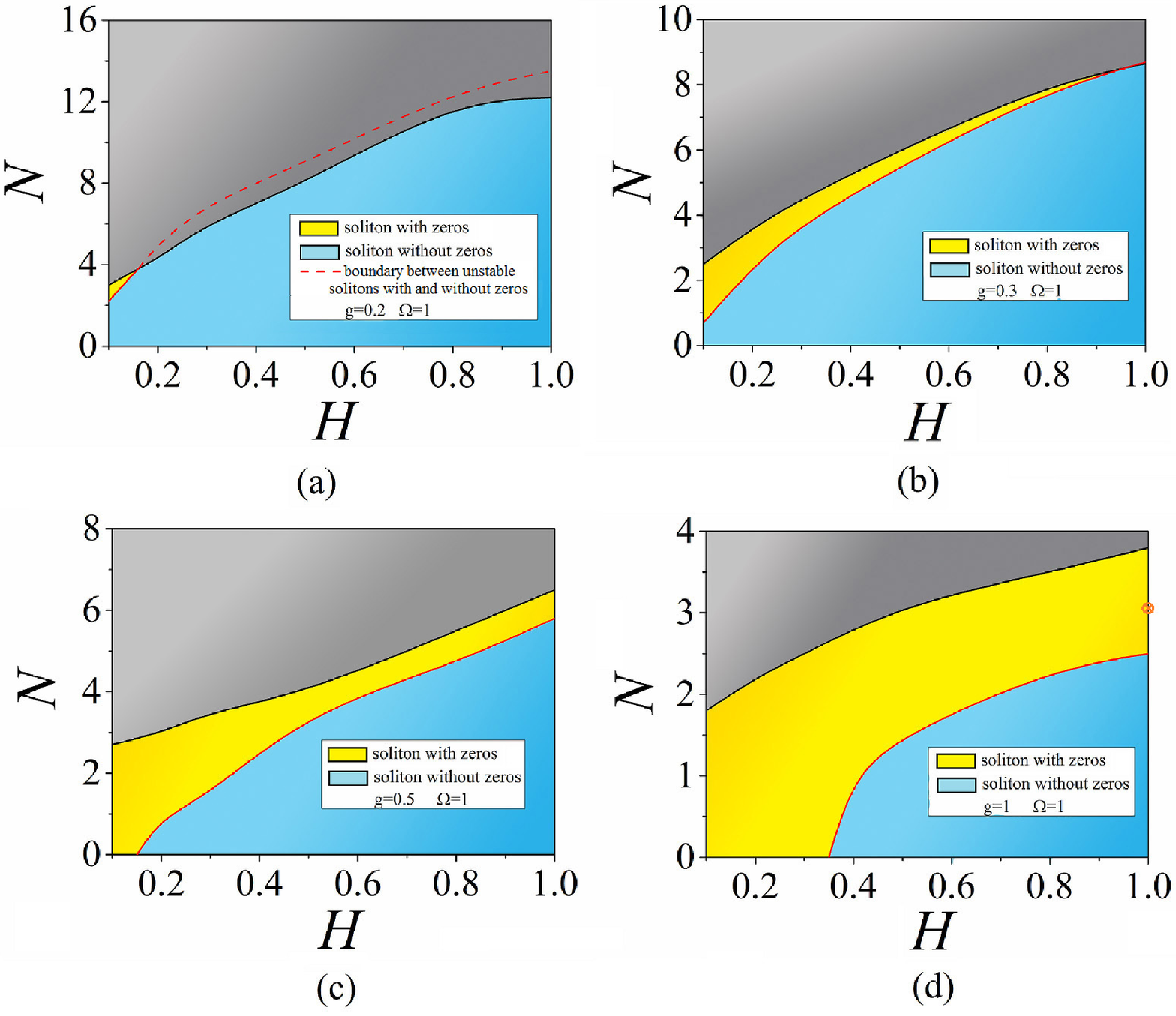}}
\caption{Stability regions for asymmetric solitons in the parametric plane
of $\left( H,N\right) $ with $(\Omega ,\protect\gamma )=(1,0.5)$. The
solitons are unstable in the gray area (upper left), being stable as modes without and
with intrinsic zero (node) in the blue (lower right) and yellow (middle) regions, respectively. In
panel (a), unstable solitons in the stripe between the black and dashed red
lines feature a node. The cross-in-circle symbol in panel (d) represents the
stable soliton shown in Fig. \protect\ref{3a}(f).}
\label{4a}
\end{figure}

\begin{figure}[tbp]
\centering{\label{fig5a} \includegraphics[scale=0.3]{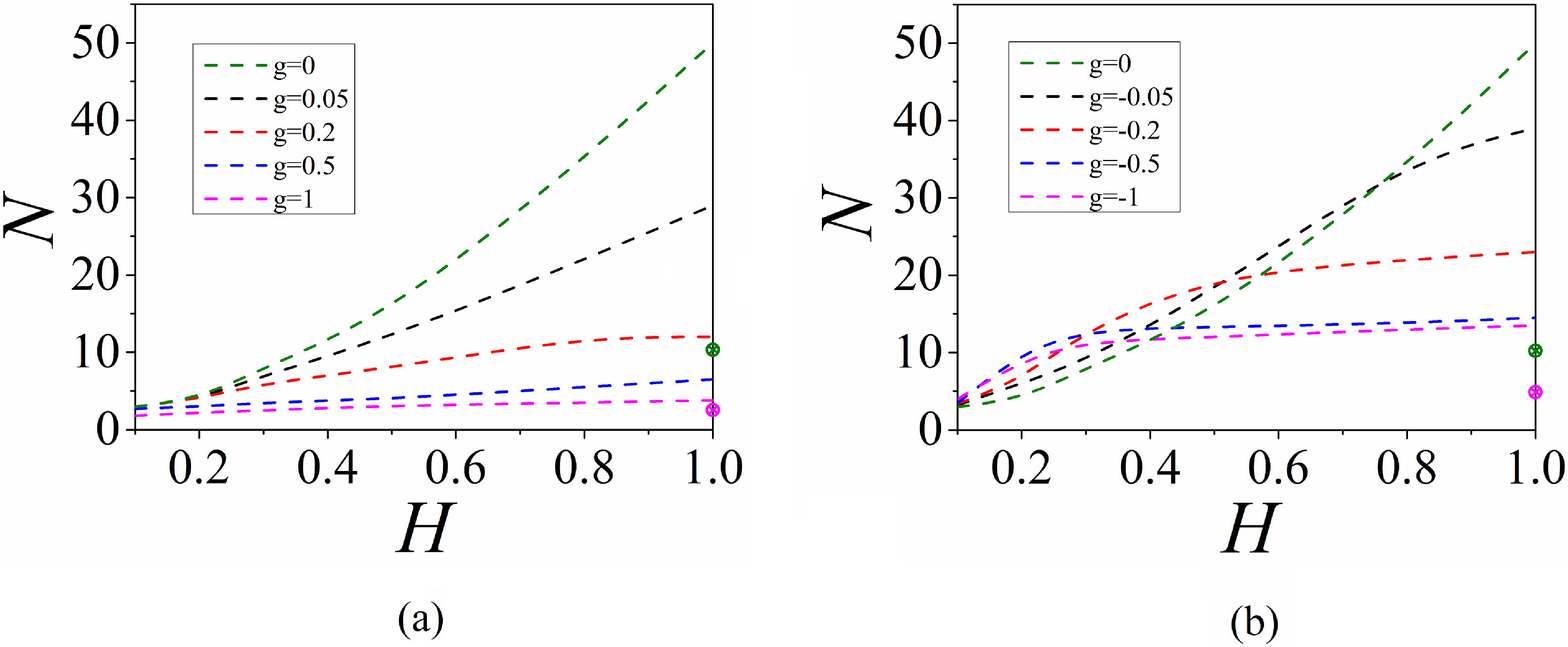}}
\caption{Asymmetric solitons are stable beneath boundaries in the plane of $%
\left( H,N\right) $, displayed in this figure for $(\Omega ,\protect\gamma %
)=(1,0.5)$, and a set of values of the self-interaction coefficient,
corresponding to attraction [$g\geq 0$ in (a)] and repulsion [$g\leq 0$ in
(b)]. The cross-in-circle symbols in panels (a) and (b) represent the stable
solitons shown in Figs. \protect\ref{3a}(e,f) and (d,e), respectively. On the section $H=1$, each line from the top to the bottom features the same order shown in the inset chart. }
\label{5a}
\end{figure}

In the case of $\Omega =1$, results for the shape of the asymmetric
fundamental solitons (nodeless or noded) and their stability are reported in
Figs. \ref{4a} and \ref{5a}, in the parameter plane of $\left( H,N\right) $,
for different values of the contact-interaction constant, $g$. Panels (a)
through (d) in Fig. \ref{4a} clearly demonstrate that the node appears in
the stable ground-state soliton at $g\approx 0.2$, and the corresponding
parameter area expands with the increase of $g$. Areas of stable and
unstable solitons are also identified in Fig. \ref{4a}. Additionally, the
stability boundaries are displayed, for a set of positive and negative
values of $g$, in Fig. \ref{5a}. It is seen that the increase of the
background magnetic field, i.e., Rabi coupling between the two components of
the binary wave function, helps to stabilize the solitons. Generally, the
increase of $|g|$ leads to destabilization, although in the case of the
contact self-repulsion, $g<0$, the situation is nearly opposite at $H<0.3$.

Note that the stability maps displayed in Figs. \ref{4a} and \ref{5a} are
plotted for $H\geq 0.1$. At smaller values of $H$, the solitons are very
broad, being distorted by boundaries of the integration domain.

Next, we report systematic results produced for families of skew-symmetric
solitons by the system without the ZS, $\Omega =0$, in Fig. \ref{6a}(a). In
this case, $H=1$ may be fixed by scaling, hence the full stability regions
are displayed in the plane of $\left( g,N\right) $, including both positive
and negative values of $g$. The plot clearly demonstrates monotonous
shrinkage of the stability area with the increase of $|g|$, which generally
resembles the trend for asymmetric solitons, observed in Figs. \ref{4a} and %
\ref{5a} in the case of $\Omega =1$.

\begin{figure}[h]
\centering{\label{fig6a} \includegraphics[scale=0.5]{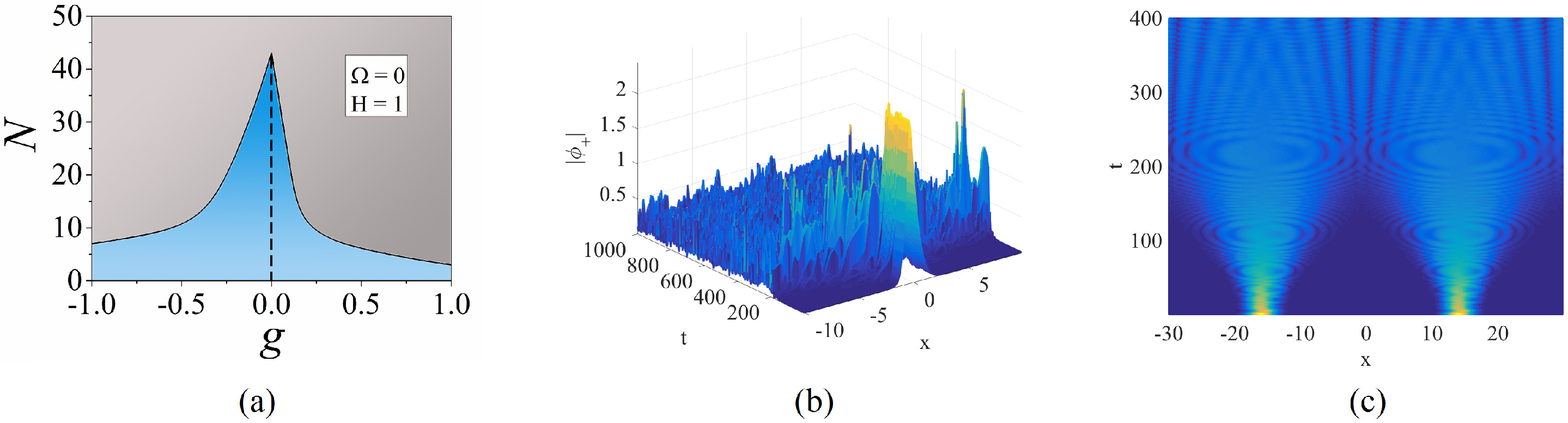}}
\caption{In panel (a), skew-symmetric fundamental solitons are stable and
unstable, respectively, in blue (bottom) and gray (top) areas in the plane of $(g,N)$, for
fixed parameters $\Omega =0$, $H=1$, and $\protect\gamma =0.5$. (b) A
typical example of the evolution of component $\protect\phi _{+}$ in an
unstable soliton, with parameters $(N,\Omega ,H,g,\protect\gamma %
)=(8,0,1,0.5,0.5)$ (the instability is similar in the other component). (c)
The simulation of the interaction between two identical stable solitons with
phase shift $\Delta \protect\varphi =0$ between them. The parameters are $%
(N,\Omega ,H,g,\protect\gamma )=(10,0,1,0,0.5)$. The interaction is similar
for other values of the phase shift, including $\Delta \protect\varphi =%
\protect\pi /2$ and $\protect\pi $.}
\label{6a}
\end{figure}

\begin{figure}[tbp]
\centering{\label{fig11a} \includegraphics[scale=0.17]{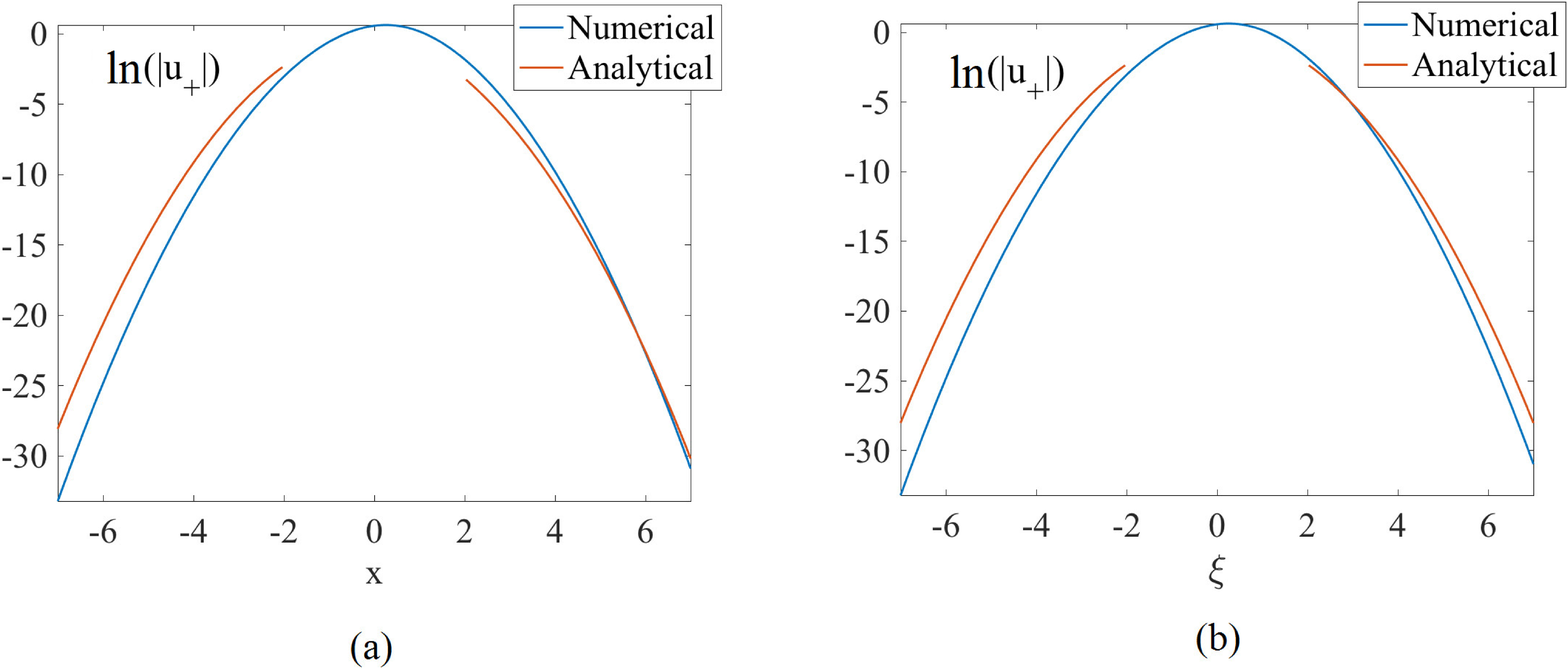}}
\caption{The comparison between the logarithmic form of the analytically
predicted Gaussian tails (red discontinuous lines) of the fundamental skew-symmetric solitons and
their numerically found counterparts (blue continuous lines). (a) The comparison for a typical
stable quiescent soliton with parameters $(N,\Omega ,H,g,\protect\gamma %
)=(10,0,1,0,0.5)$, the analytical prediction given by Eq. (\protect\ref%
{asympt}). (b) The comparison for a typical moving soliton, for $(N,\Omega
,H,g,\protect\gamma ,v)=(10,0,1,0,0.5,0.1)$, with the analytical prediction
provided by Eq. (\protect\ref{asympt2}). The comparison is displayed only
for component $u_{+}$, the pictures for $u_{-}$ being mirror images of the
present ones.}
\label{11a}
\end{figure}

Tails of the numerically found solitons indeed feature a Gaussian shape, as
predicted by Eq. (\ref{asympt}). The analytical prediction for the tails of
a skew-symmetric soliton is compared to its numerically generated
counterparts in Fig. \ref{11a}(a), which demonstrates that the simple
analytical approximation provides reasonable accuracy.

As concerns unstable solitons, direct simulations demonstrate that they\ are
quickly destroyed, being replaced by a finite-amplitude turbulent state, as
shown in Fig. \ref{6a}(b) (the unstable evolution is similar in the case of $%
\Omega =1$). Further, creating a pair of spatially separated solitons, it is
natural to expect that the nonlocal interaction, mediated by the microwave
field, will also induce a long-range interaction between non-overlapping
solitons. This is indeed observed in Fig. \ref{6a}(c), which demonstrates
that the long-range interaction between the solitons destroys them, leading
to creation of a quasi-linear interference pattern (which may be affected by
reflections from edges of the integration domain, if the latter is not broad
enough). A detailed study of interactions between solitons in the rpesent
system may be a subject for a separate work.

\subsection{Dipole solitons}

\begin{figure}[h]
\centering{\label{fig7a} \includegraphics[scale=0.32]{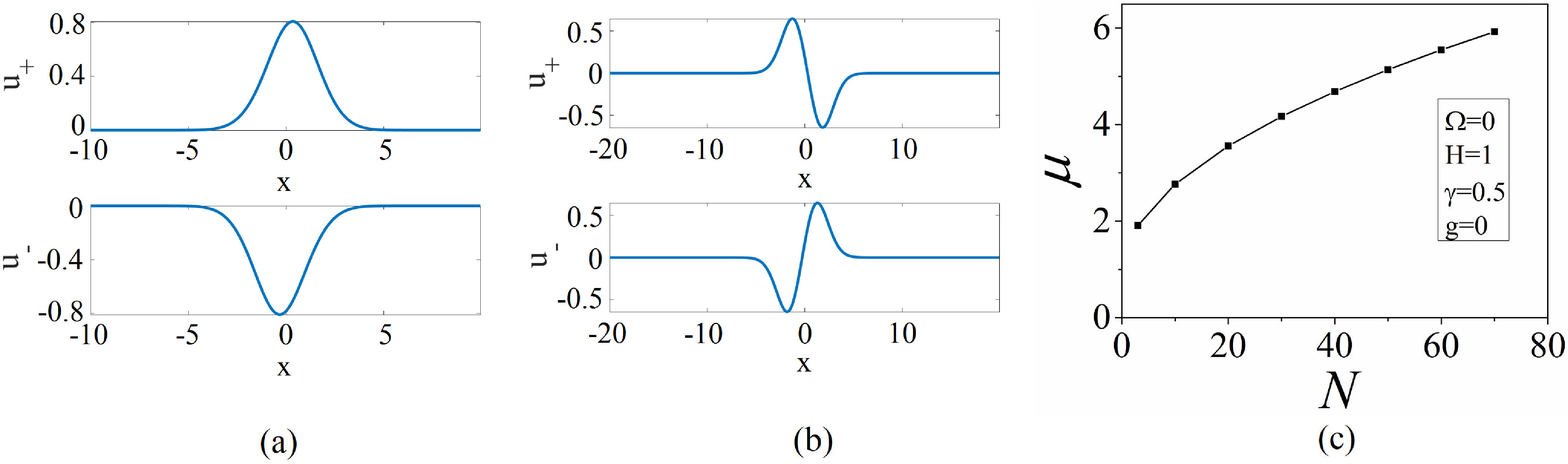}}
\caption{(a) and (b): Examples of stable skew-symmetric fundamental and
dipole solitons in the absence of the Zeeman splitting and contact
nonlinearity, i.e., with $\Omega =0$, $H=1$, $g=0$ and $\protect\gamma =0.5$%
, and equal norms, $N=3$. The respective chemical potentials are $\protect%
\mu _{\mathrm{fund}}=1.47$ and $\protect\mu _{\mathrm{dip}}=1.91$. (c)
Dependence $N(\protect\mu )$ for the family of stable skew-symmetric dipole
solitons.}
\label{7a}
\end{figure}

\begin{figure}[h]
\centering{\label{fig8a} \includegraphics[scale=0.32]{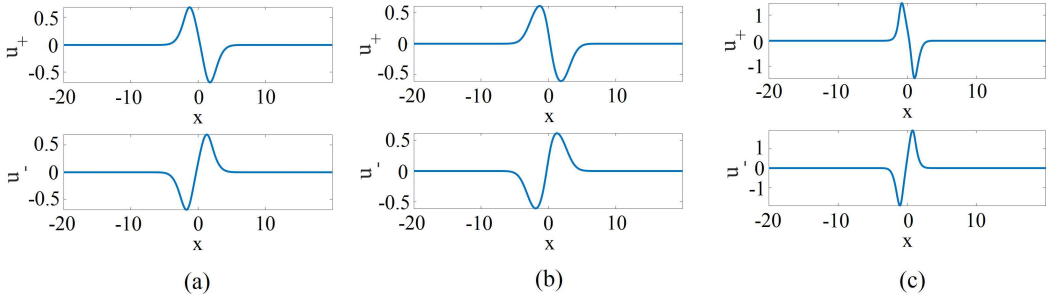}}
\caption{(a,b) Typical examples of stable skew-symmetric dipole solitons in
the system with $(\Omega ,H,\protect\gamma )=(0,1,0.5)$, which includes the
contact self-attraction and repulsion, respectively, with $g=\pm 0.5$. Both
solitons have equal norms, $N=3$, the chemical potentials being $\protect\mu %
_{a}=1.71$ and $\protect\mu _{b}=2.08$, respectively. (c) A stable
skew-asymmetric dipole soliton with parameters $(N,\Omega ,H,g,\protect%
\gamma )=(10,1,1,0.3,0.5)$. Note that a fundamental soliton with the same
parameters as in (a) is unstable, as per Fig. \protect\ref{4a}(b).}
\label{8a}
\end{figure}

The above consideration addressed solely fundamental (ground-state)
solitons. The fact that they are supported by the effective trapping
potential in Eq. (\ref{II}) suggests a possibility to look for higher-order
solitons, that may correspond to excited bound states in the trapping
potential (such stable states were not reported in related works \cite%
{Jieli-SOC}, \cite{we}, and \cite{HS}; in the two latter papers, excited
states were addressed, but they all were found to be unstable). The present
system readily creates \emph{stable} excited states in the form of dipole
(spatially odd) modes, which, in the case of $\Omega =0$, satisfy the
skew-symmetry condition in the form of Eq. (\ref{skew}) with the top sign
(on the contrary to the bottom sign corresponding to the fundamental
states), as shown in Figs. \ref{7a}(b) and \ref{8a}(a,b) for the system
without and with the contact interaction, \textit{viz}., $g=0$ and $g=\pm 0.5
$, respectively. Further, Figs. \ref{7a}(a) and (b) compare the fundamental
soliton and its dipole counterpart at identical values of the parameters,
including the total norm (but, naturally, with different chemical
potentials, $\mu _{\mathrm{fund}}=1.47$ and $\mu _{\mathrm{dip}}=1.91$), in
the system with $g=0$. The results are summarized in Fig. \ref{7a}(c), which
displays the $\mu (N)$ dependence for the family of stable skew-symmetric
dipole solitons. Note that the family satisfies the anti-VK stability
criterion, $d\mu /dN>0$, which is relevant in this case, as the contact
self-attraction is absent.

Next, Fig. \ref{8a}(c) displays an example of a stable dipole mode in which
the skew symmetry is broken by the ZS, with $\Omega =1$. It is worthy to
mention that, for the same parameters as used in the latter case, i.e., $%
(N,\Omega ,H,g,\gamma )=(10,1,1,0.3,0.5)$, the fundamental soliton falls in
the unstable (gray) area in Fig. \ref{4a}(b), even if its chemical
potential, $\mu _{\mathrm{fund}}=0.73$, is much smaller than $\mu _{\mathrm{%
dip}}=2.14$ of the stable dipole soliton with the same norm. Thus, the
dipole solitons may be \emph{more stable} than their fundamental
counterparts, at the same parameters and with the same norm. To the best of
our knowledge, no previous model produced a larger stability area for dipole
solitons than for fundamental ones (in a specific model with spatially
modulated nonlinearity, introduced in Ref. \cite{Barcelona}, the fundamental
and dipolar solitons are completely stable in their entire existence area,
while instability appears in higher-order excited states).

\subsection{Solitons' mobility}

Generation of moving solitons from the quiescent ones considered above is a
nontrivial problem, as Eq. (\ref{basiceq}), obviously, is not Galilean
invariant. To this end, we rewrite the equations in the reference frame
moving with velocity $v$, cf. Ref. \cite{Ben Li}:

\begin{eqnarray}
&&i\partial _{t}\phi _{+}-iv\partial _{\xi }\phi _{+}=\partial _{\xi }\phi
_{-}-\Omega \phi _{+}-H\phi _{-}-g|\phi _{+}|^{2}\phi _{+}+\frac{\gamma }{2}{%
\phi }_{-}\int_{-\infty }^{+\infty }|\xi -\xi ^{\prime }|{\phi }_{-}^{\ast
}(\xi ^{\prime }){\phi }_{+}(\xi ^{\prime })d\xi ^{\prime },  \notag \\
&&i\partial _{t}\phi _{-}-iv\partial _{\xi }\phi _{-}=-\partial _{\xi }\phi
_{+}+\Omega \phi _{-}-H\phi _{+}-g|\phi _{-}|^{2}\phi _{-}+\frac{\gamma }{2}{%
\phi }_{+}\int_{-\infty }^{+\infty }|\xi -\xi ^{\prime }|{\phi }_{-}(\xi
^{\prime }){\phi }_{+}^{\ast }(\xi ^{\prime })d\xi ^{\prime },
\label{mobilityeq}
\end{eqnarray}%
where $\xi \equiv x-vt$ is the moving coordinate. In this reference frame,
stationary solutions are looked for as $\phi _{\pm }=\exp \left( -i\mu
t\right) u_{\pm }(\xi )$, with complex stationary wave function $u_{\pm }$
satisfying equations%
\begin{eqnarray}
&&\mu u_{+}-iv\partial _{\xi }u_{+}=\partial _{\xi }u_{-}-\Omega
u_{+}-Hu_{-}-g|u_{+}|^{2}u_{+}+\frac{\gamma }{2}{u}_{-}\int_{-\infty
}^{+\infty }|\xi -\xi ^{\prime }|{u}_{-}^{\ast }(\xi ^{\prime }){u}_{+}(\xi
^{\prime })d\xi ^{\prime },  \notag \\
&&\mu u_{-}-iv\partial _{\xi }u_{-}=-\partial _{\xi }u_{+}+\Omega
u_{-}-Hu_{+}-g|u_{-}|^{2}u_{-}+\frac{\gamma }{2}{u}_{+}\int_{-\infty
}^{+\infty }|\xi -\xi ^{\prime }|{u}_{-}(\xi ^{\prime }){u}_{+}^{\ast }(\xi
^{\prime })d\xi ^{\prime }.  \label{uu}
\end{eqnarray}%
Note that, in the absence of the ZS ($\Omega =0$), solutions of Eq. (\ref{uu}%
) satisfy the skew-symmetry condition generalized for the complex wave
functions: $u_{+}(\xi )=\pm u_{-}^{\ast }(-\xi )$, cf. Eq. (\ref{skew}).

In the asymptotic area of $|\xi |\rightarrow \infty $, Eq. (\ref{uu}) takes
the form of%
\begin{eqnarray}
&&\mu u_{+}-iv\partial _{\xi }u_{+}=\partial _{\xi }u_{-}-\Omega
u_{+}-Hu_{-}+\frac{\gamma }{2}\tilde{I}|\xi |{u}_{-},  \notag \\
&&\mu u_{-}-iv\partial _{\xi }u_{-}=-\partial _{\xi }u_{+}+\Omega
u_{-}-Hu_{+}+\frac{\gamma }{2}\tilde{I}^{\ast }|\xi |{u}_{+},  \label{IIII}
\end{eqnarray}%
where
\begin{equation}
\tilde{I}\equiv \int_{-\infty }^{+\infty }u_{-}^{\ast }(\xi )u_{+}(\xi )d\xi
,  \label{tilde}
\end{equation}%
cf. Eq. (\ref{II}). The analysis of Eq. (\ref{IIII}) yields, in the lowest
approximation, the Gaussian asymptotic profile of the moving soliton,
\begin{equation}
\left\{ u_{-}(x),u_{+}(x)\right\} \sim \exp \left\{ -\frac{\sqrt{\left[
\mathrm{Re}(\tilde{I})\right] ^{2}-v^{2}\left\vert \tilde{I}\right\vert ^{2}}%
-i\mathrm{Im}(\tilde{I})\mathrm{sgn}\xi }{4\left( 1-v^{2}\right) }\gamma \xi
^{2}\right\} ,  \label{asympt2}
\end{equation}%
cf. Eq. (\ref{asympt}). The Gaussian tails of a stable skew-symmetric moving
soliton are compared to their analytical counterparts, predicted by Eq. (\ref%
{asympt2}), in Fig. \ref{11a}(b). Similar to their quiescent counterparts
[cf. Fig. \ref{11a})], the moving solitons are approximated reasonably well
by the Gaussian ansatz for the tails.

As it follows from Eq. (\ref{asympt2}), the moving solitons may exist for
velocities falling below a certain critical value, which is determined by an
implicit condition%
\begin{equation}
v^{2}<v_{\max }^{2}=\left[ \mathrm{Re}(\tilde{I})\right] ^{2}/\left\vert
\tilde{I}\right\vert ^{2}<1.  \label{max}
\end{equation}%
Obviously, Eq. (\ref{max}), which is produced by the consideration of the
asymptotic form of the solitons' tails, is an upper limit, but not a
sufficient condition, for the existence of moving solitons. Indeed,
numerical findings demonstrate that largest velocities, up to which the
solitons can be found, are smaller than the value given by Eq. (\ref{max}).

\begin{figure}[tbp]
\centering{\label{fig9a} \includegraphics[scale=0.15]{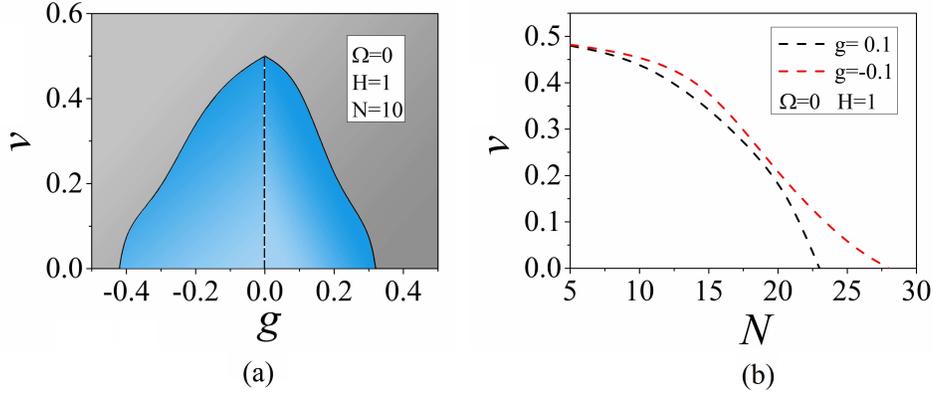}}
\caption{Stability areas for skew-symmetric solitons moving with velocity $v$
[blue in panel (a), and beneath the dashed boundaries in (b)], in parameter
planes $\left( g,v\right) $ and $\left( N,v\right) $. In panel(b), $g=-0.1$ corresponds to the red dashed line (upper). Values of other
parameters are indicated in the panels.}
\label{9a}
\end{figure}

Numerical results for moving skew-symmetric solitons are summarized in the
form of stability regions in the planes of $\left( g,v\right) $ and $\left(
N,v\right) $ shown in Fig. \ref{9a} for $(\Omega ,H)=(0,1)$. The figure
demonstrates that the stability regions shrink with the increase of both $|g|
$ and $N$, the largest velocity admitting the stability being attained in
the absence of the contact nonlinearity, $g=0$. Note that the maximum
velocity in Fig. \ref{9a}(a) is $\approx 0.5$, being, indeed, smaller than
the respective value $v_{\max }\approx 0.84$, which is predicted by the
upper limit given by Eq. (\ref{max}).

\begin{figure}[tbp]
\centering{\label{fig10a} \includegraphics[scale=0.3]{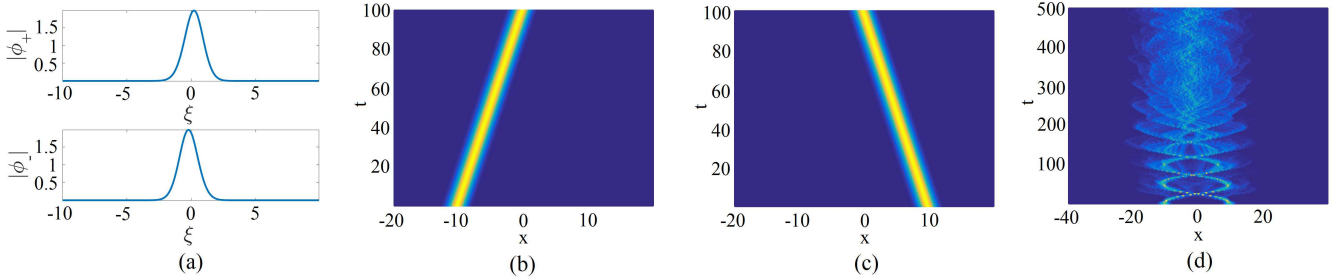}}
\caption{(a) A moving skew-symmetric soliton located near the stability
boundary, for $(N,H,\Omega ,\protect\gamma ,g,v)=(10,1,0,0.5,0,0.5)$. (b,c)
Density plots for solitons moving with velocities $v=\pm 0.1$ are displayed
in the quiescent reference frame, for parameters $(N,H,\Omega ,\protect%
\gamma ,g)=(10,1,0,0.5,0)$. (d) Collision between solitons moving with
opposite velocities, $v=\pm 0.1$.}
\label{10a}
\end{figure}

A typical example of a skew-symmetric moving soliton (with velocity $v=0.5$)
near the stability boundary is shown in Fig. \ref{10a}(a). Further, the
motion of solitons with velocities $v=\pm 0.1$ is illustrated by the density
plots displayed in Figs. \ref{10a}(b,c). Finally, the availability of the
solitons moving in opposite directions makes it possible to simulate
collisions between them. A typical outcome of collisions is displayed in
Fig. \ref{10a}(d). Due to the effective nonlocality of the
microwave-mediated coupling, the solitons commence interacting before coming
in contact, similar to what is shown above in Fig. \ref{6a}(c). After
several collisions, the two solitons merge into a single one, at $t>220$.

\section{Conclusion}

This work demonstrates that the concept of gap solitons in two-component
BECs with the kinetic energy negligible in comparison with the SOC
(spin-orbit-coupling) and ZS (Zeeman-splitting) terms in the Hamiltonian may
be applied to the system in which two atomic states are resonantly coupled
by the microwave radiation, the feedback of the two-component atomic wave
function on the microwave field being represented by the Poisson's equation
(which is solved by means of the Green's function). The nonlocal interaction
between the components, mediated by the radiation, drastically changes the
concept of the linear spectrum, adding to it an effective nonlinear trapping
potential $\sim |x|$,  thus making the system non-linearizable. As a result,
the system may create families of both gap and embedded solitons, a
considerable part of which is stable (the gap tends to remain empty while
embedded solitons exist). In the case when the system includes repulsive\
(or zero) contact nonlinearity, the stability of soliton families obeys the
anti-Vakhitov-Kolokolov criterion. The two-component solitons feature exact
or approximate (broken) skew-symmetric shapes. In addition to ground-state
fundamental solitons, which may feature a node in their shape, the system
supports dipole solitons, whose stability area is broader than for their
fundamental counterparts. The stability area is identified too for moving
solitons, being limited by a largest values of the velocity, and collisions
between moving solitons lead to merger into a single one, via a complex
interaction.

A challenging problem for further analysis is the consideration of the
two-dimensional version of the present system. In particular, the Green's
function for the one-dimensional Poisson's equation should be replaced by
its two-dimensional (logarithmic) counterpart \cite{Jieli-SOC}.

\section*{Acknowledgment}

This work is supported, in part, by the Israel Science Foundation through
grant No. 1287/17, NNSFC (China) through Grants No. 11874112 and 11575063, and by the State Scholarship Fund of China Scholarship council through File No. 201808440001. Z. Fan appreciates technical assistance provided by Ms. Shiyue Liu (Chinese University of Hong Kong).

\end{document}